      [1999/12/01 v1.4c Il Nuovo Cimento]
\documentclass{cimento}

\usepackage{epsfig}
\usepackage{amsmath}
\usepackage{latexsym}
\usepackage{amssymb}
\usepackage{dsfont}
\usepackage{color}
\usepackage{multirow}

\newcommand{\ud}{\mathrm{d}}

\title{Hadron tomography through Wigner distributions}
\author{C.~Lorc\'e\from{ins:upsud}
        \atque
B.~Pasquini\from{ins:pavia}}
\instlist{\inst{ins:upsud} IPN and LPT, Universit\'e Paris-Sud, 91405 Orsay, France
  \inst{ins:pavia} Dipartimento di Fisica Nucleare e Teorica, Universit\`a degli Studi di Pavia, and INFN, Sezione di Pavia, I-27100 Pavia, Italy}
\PACSes{\PACSit{12.38.-t}{Quantum chromodynamics}\PACSit{12.39.-x}{Phenomenological quark models}\PACSit{14.20.Dh}{Protons and neutrons}}
\begin{document}

\maketitle

\begin{abstract}
We study the Wigner functions of the nucleon which provide multidimensional images of the quark distributions in phase space and combine in a single picture all the information contained in the generalized parton distributions (GPDs) and the transverse-momentum dependent parton distributions (TMDs). In particular, we present results for the distribution of unpolarized quarks in a longitudinally polarized nucleon obtained in a light-cone constituent quark model. Treating the Wigner distribution as it was a classical distribution, we obtain the quark orbital angular momentum and compare it with alternative definitions given in terms of the GPDs and the TMDs.
\end{abstract}

\section{Introduction}
\label{section-1}

Quark Wigner distributions provide joint position-and-momentum (or phase-space) distributions, encoding in a unified picture  the information  obtained from transverse-momentum dependent parton distributions (TMDs) and generalized parton distributions (GPDs) in impact-parameter space. The concept of Wigner distributions in QCD for quarks and gluons was first explored in refs.~\cite{Ji:2003ak,Belitsky:2003nz}. Neglecting relativistic effects, the authors introduced six-dimensional (three position and three momentum coordinates) Wigner distributions. In a recent work~\cite{Lorce:2011kd}, we used the connection between Wigner distributions and generalized transverse-momentum dependent parton distributions (GTMDs)~\cite{Meissner:2009ww} to study five-dimensional distributions (two position and three momentum coordinates) in the infinite-momentum frame providing a picture consistent with special relativity. However, it is well known that the quantum-mechanical phase-space distributions do not have a density interpretation because of the uncertainty principle. Accordingly, Wigner distributions are not positively defined. Nevertheless, the physics of the Wigner distributions is very rich and one can select certain situations where a semiclassical interpretation is still possible.

The purpose of this contribution is to investigate the phenomenology of the quark Wigner distributions. As a matter of fact, since it is not known how to access these distributions directly from experiments, phenomenological models are very powerful in this context. Collecting the information that one can learn from quark models which were built up on the basis of available experimental information on GPDs and TMDs, one can hope to reconstruct a faithful description of the physics of the Wigner distributions. To this aim we will rely on models for the light-cone wave functions which have already been used for the description of the GPDs~\cite{Boffi:2007yc,Lorce:2011dv}, the TMDs~\cite{Lorce:2011dv,Pasquini:2008ax,Boffi:2009sh,Pasquini:2010af,Lorce:2011zt} and electroweak properties of the nucleon~\cite{Lorce:2011dv,Lorce:2006nq,Lorce:2007as,Lorce:2007fa,Pasquini:2007iz}.

\section{Wigner distributions}

Similarly to refs.~\cite{Ji:2003ak,Belitsky:2003nz}, we define the Hermitian Wigner operators for quarks at a fixed light-cone time $y^+=0$ as follows
\begin{equation}\label{wigner-operator}
\widehat W^{[\Gamma]}(\vec b_\perp,\vec k_\perp,x)\equiv\frac{1}{2}\int\frac{\ud z^-\,\ud^2z_\perp}{(2\pi)^3}\,e^{i(xp^+z^--\vec k_\perp\cdot\vec z_\perp)}\,\overline{\psi}(y-\tfrac{z}{2})\Gamma\mathcal W\,\psi(y+\tfrac{z}{2})\big|_{z^+=0},
\end{equation}
with $y^\mu=[0,0,\vec b_\perp]$, $p^+$ the average nucleon longitudinal momentum and $x=k^+/p^+$ the average fraction of nucleon longitudinal momentum carried by the active quark. The superscript $\Gamma$ stands for any twist-two Dirac operator $\Gamma=\gamma^+,\gamma^+\gamma_5,i\sigma^{j+}\gamma_5$ with $j=1,2$. A Wilson line $\mathcal W\equiv\mathcal W(y-\tfrac{z}{2},y+\tfrac{z}{2}|n)$ ensures the color gauge invariance of the Wigner operator. In the following we will focus on the quark contribution, ignoring the gauge-field degrees of freedom and therefore reducing the gauge link $\mathcal W$ to the identity. 

We define the Wigner distributions $\rho^{[\Gamma]}(\vec b_\perp,\vec k_\perp,x,\vec S)$ in terms of the matrix elements of the Wigner operators \eqref{wigner-operator} sandwiched between nucleon states with polarization $\vec S$
\begin{equation}\label{wigner}
\rho^{[\Gamma]}(\vec b_\perp,\vec k_\perp,x,\vec S)\equiv\int\frac{\ud^2\Delta_\perp}{(2\pi)^2}\,\langle p^+,\tfrac{\vec\Delta_\perp}{2},\vec S|\widehat W^{[\Gamma]}(\vec b_\perp,\vec k_\perp,x)|p^+,-\tfrac{\vec\Delta_\perp}{2},\vec S\rangle.
\end{equation}
As outlined in ref.~\cite{Lorce:2011kd}, such matrix elements can be interpreted as two-dimensional Fourier transforms of the GTMDs in the impact-parameter space. Although the GTMDs are in general complex-valued functions, their two-dimensional Fourier transforms are always real-valued functions, in accordance with their interpretation as phase-space distributions. We note that, like in the usual quantum-mechanical Wigner distributions, $\vec b_\perp$ and $\vec k_\perp$ are not Fourier conjugate variables. However, they are subjected to Heisenberg's uncertainty principle because the corresponding quantum-mechanical operators do not commute $[\hat{\vec b}_\perp,\hat{\vec k}_\perp]\neq 0$. As a consequence, the Wigner functions can not have a strict probabilistic interpretation. There are in total 16 Wigner functions at twist-two level, corresponding to all the 16 possible configurations of nucleon and quark polarizations. Here we will discuss only one particular case, namely the distortion in the distribution of unpolarized quarks due to the longitudinal polarization of the nucleon $\rho_{LU}=\rho^{[\gamma^+]}(\vec b_\perp,\vec k_\perp,x,+\vec e_z)-\rho^{[\gamma^+]}(\vec b_\perp,\vec k_\perp,x,-\vec e_z)$ which has a close connection with the quark orbital angular momentum (OAM). Other configurations for the quark and nucleon polarizations can be found in ref.~\cite{Lorce:2011kd}.
\begin{figure}[th!]
	\centering
		\includegraphics[width=.49\textwidth]{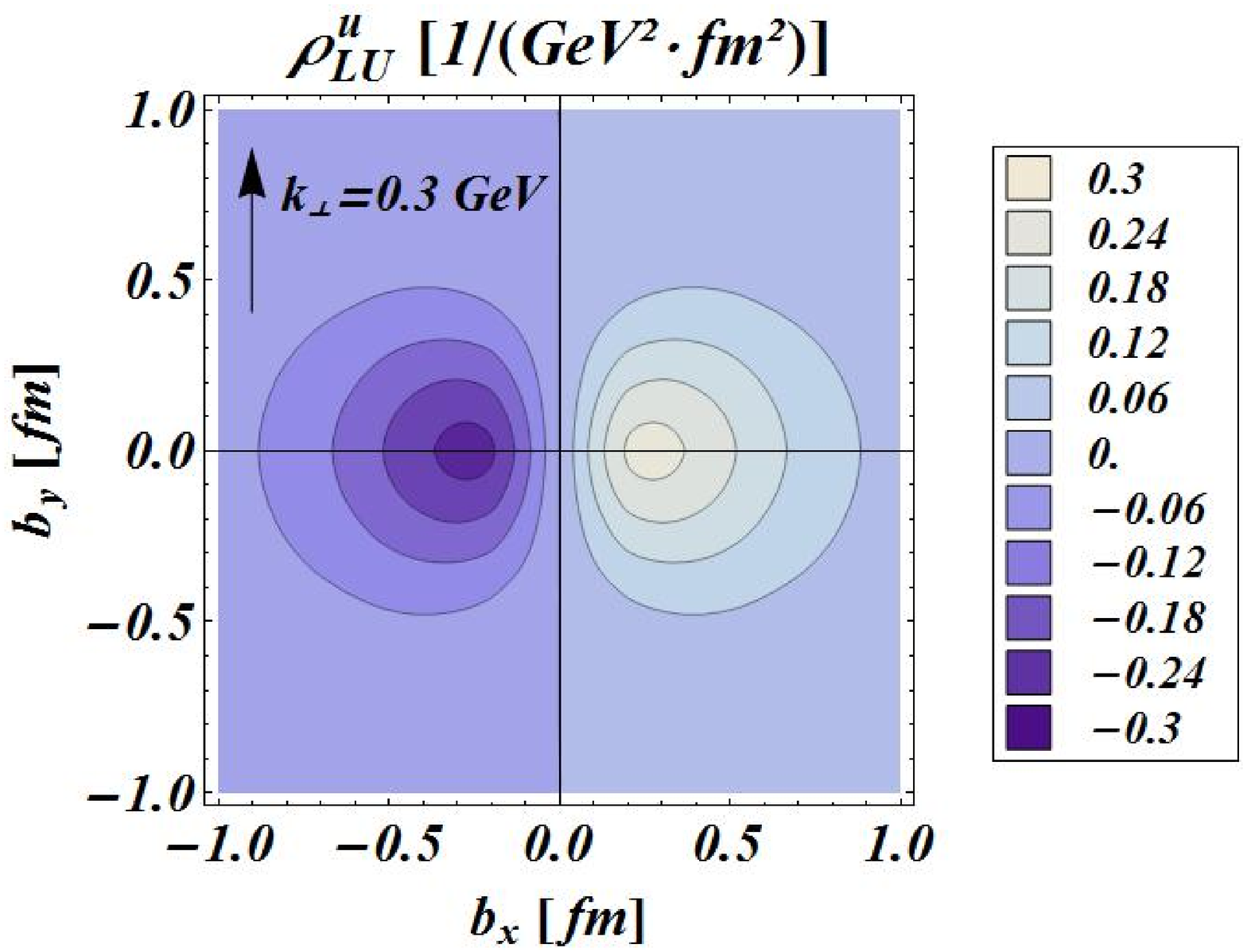}
		\includegraphics[width=.49\textwidth]{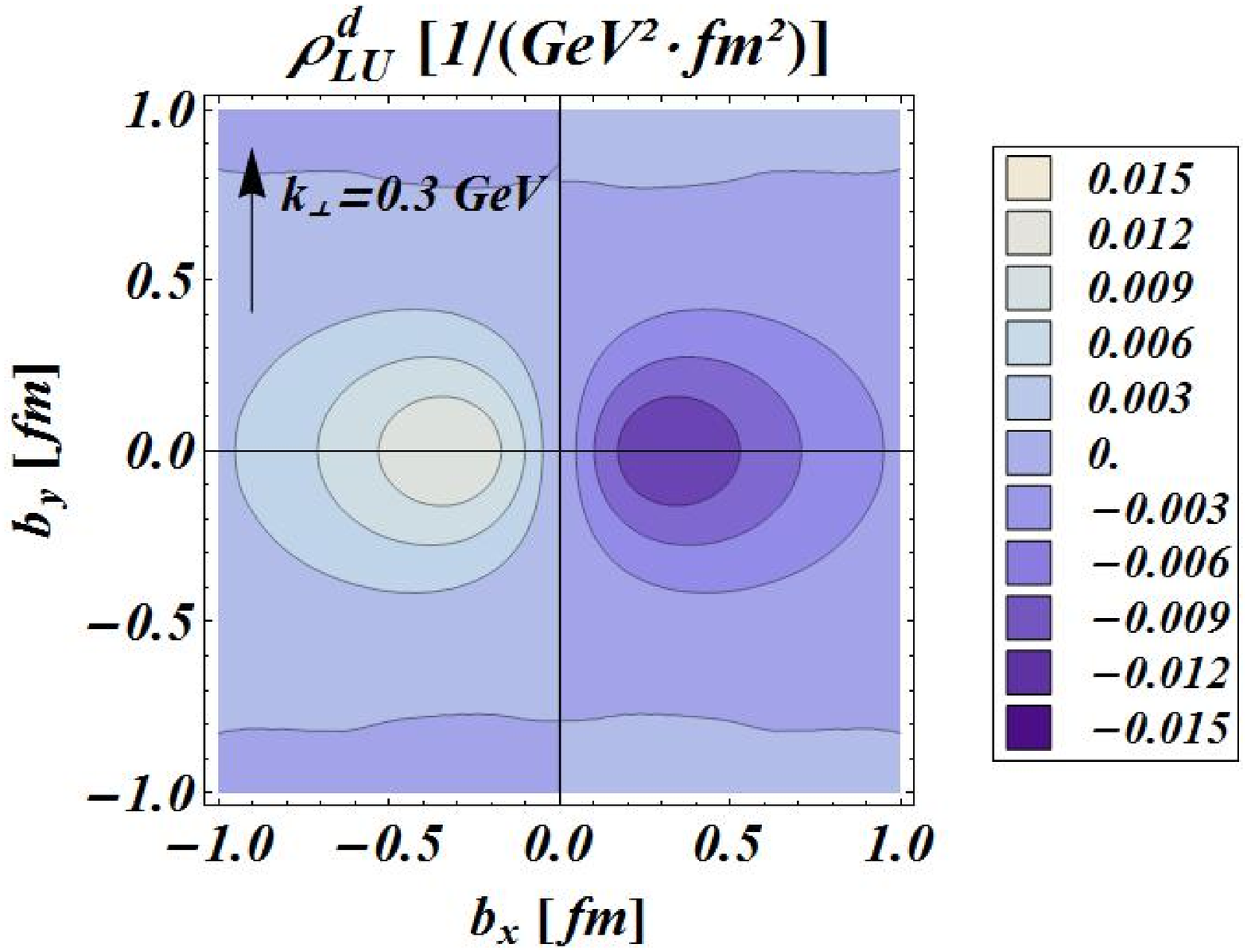}
		\includegraphics[width=.49\textwidth]{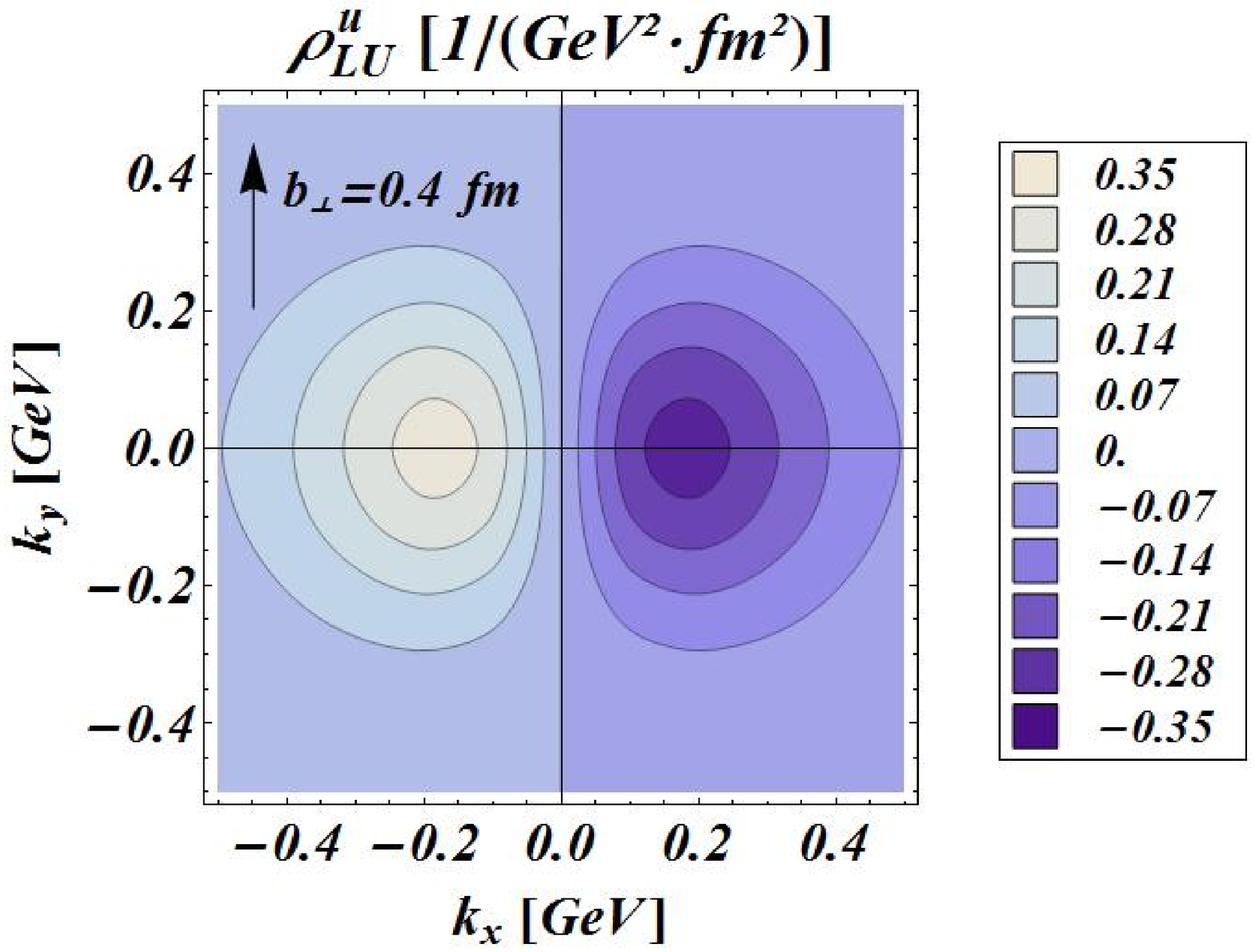}
		\includegraphics[width=.49\textwidth]{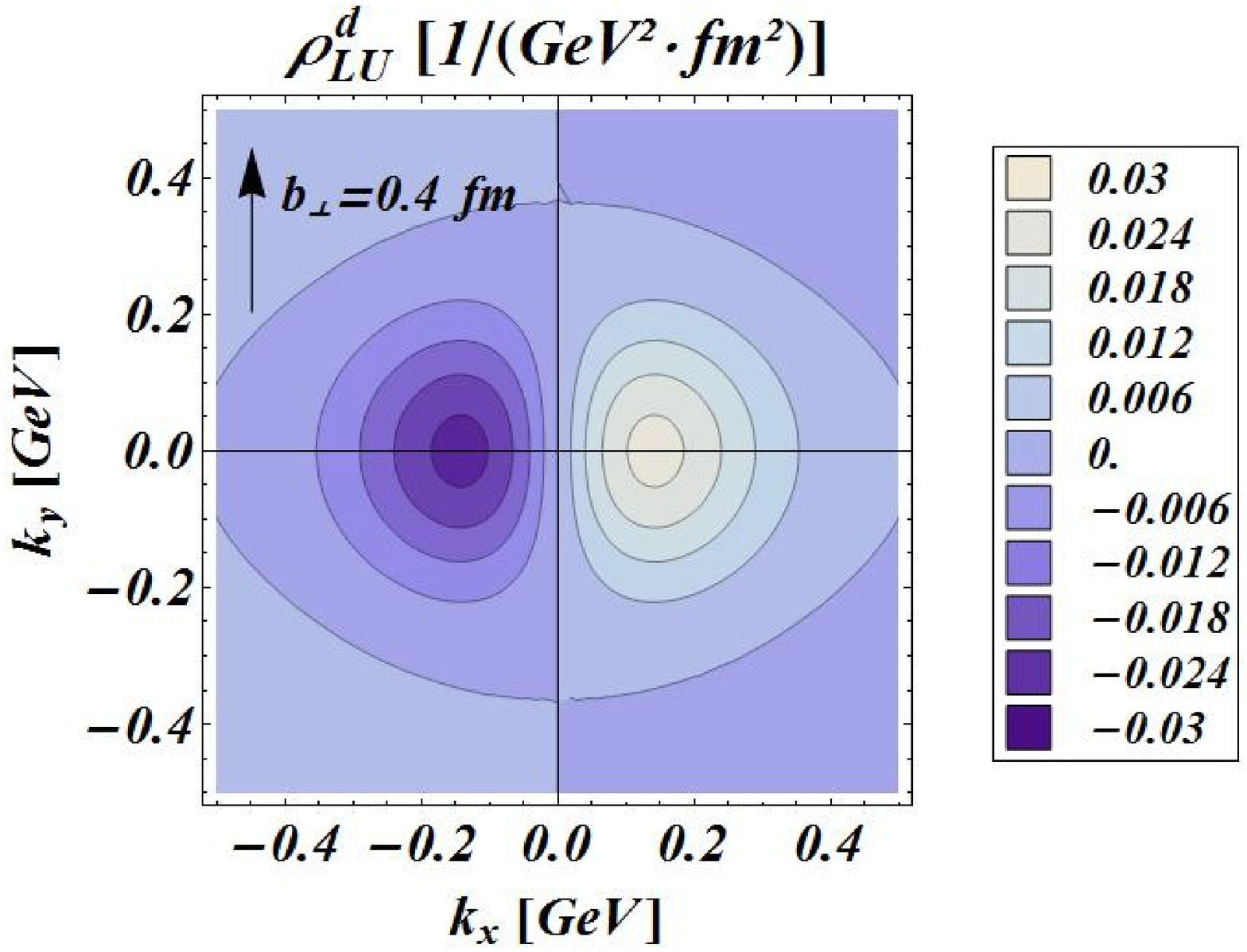}
	\caption{\footnotesize{The distortions of the $x$-integrated Wigner distributions of unpolarized quarks due to the spin of the proton (pointing out of the plane) in a LCCQM. Upper panels: distortions in impact-parameter space with fixed transverse momentum $\vec k_\perp=k_\perp\,\vec e_y$ and $k_\perp=0.3$ GeV. Lower panels: distortions in transverse-momentum space with fixed impact parameter $\vec b_\perp=b_\perp\,\vec e_y$ and $b_\perp=0.4$ fm. The left (right) panels show the results for $u$ ($d$) quarks.}}\label{fig1}
\end{figure}

In fig.~\ref{fig1}, the upper panels show the distortions in impact-parameter space for $u$ (left panels) and $d$ (right panels) quarks with fixed transverse momentum $\vec k_\perp=k_\perp\,\vec e_y$ and $k_\perp=0.3$ GeV, while the lower panels give the corresponding distortions in the transverse-momentum space with fixed impact parameter $\vec b_\perp=b_\perp\,\vec e_y$ and $b_\perp=0.4$ fm. We observe a clear dipole structure in both these distributions which indicates that the (quasi-)probability for finding the quark orbiting clockwise is not the same as for the quark orbiting anticlockwise, leading in average to a nonvanishing OAM. We learn from these figures that the OAM of $u$ quarks tends be aligned with the nucleon spin, while the OAM of $d$ quarks tends be antialigned with the nucleon spin. In particular, we notice that the distortion induced by the quark OAM is stronger in the central region of the phase space ($k_\perp\ll$ and $b_\perp\ll$), for both $u$ and $d$ quarks. The distortion in the $\vec b_\perp$-space (see upper panels of fig.~\ref{fig1}) is more extended for $d$ quarks than for $u$ quarks, whereas the opposite behavior is found for the distortion in the $\vec k_\perp$-space (see lower panels of fig.~\ref{fig1}). In the case of $d$ quarks, we also observe a sign change of the distributions in the outer regions of phase space ($k_\perp\gg$ and $b_\perp\gg$) which corresponds to a flip of the local net quark OAM. Note that such a direct access to quark OAM is not possible with GPDs and TMDs since none of them describe at leading twist the distortion in the distribution of unpolarized quarks due to the longitudinal polarization of the nucleon. 
%The reason is that 
This is because one needs the correlation between $\vec b_\perp$ and $\vec k_\perp$ which is lost by integrating over $\vec b_\perp$ or $\vec k_\perp$.

\section{Quark orbital angular momentum}

If one neglects gauge-field degrees of freedom, the quark OAM operator for a given flavor $q$ can be unambiguously defined as
\begin{equation}\label{OAMop1}
\widehat L^q_z\equiv-\frac{i}{2}\int\ud^3r\,\overline\psi^q\gamma^+\left(\vec r\times\!\stackrel{\leftrightarrow}{\nabla}_r\right)_z\psi^q,
\end{equation}
where normal ordering is understood. We define the quark OAM $\ell_z^q$ as the following matrix element of the quark OAM operator
\begin{equation}\label{oam}
\ell_z^q=\int\frac{\ud\Delta^+}{2P^+}\,\frac{\ud^2\Delta_\perp}{(2\pi)^3}\,\langle p',+|\widehat L^q_z|p,+\rangle,
\end{equation}
where the incoming and outgoing momenta of the nucleon are given by $p=P-\frac{1}{2}\Delta$ and $p'=P+\frac{1}{2}\Delta$, respectively. Since the nucleon state $|p,\Lambda\rangle$ in eq.~(\ref{oam}) is normalized as $\langle p',+|p,+\rangle=2P^+\,\delta(\Delta^+)\,(2\pi)^3\,\delta^{(2)}(\vec\Delta_\perp)$, the forward limit $p'=p$ has to be treated with care. Integrating over $\Delta^+$ and $\vec\Delta_\perp$ leads to consistent expressions while avoiding the use of normalizable wave packets or infinite normalization factors. It turns out that this definition of quark OAM coincides with treating the Wigner functions as if they were classical distributions, and obtaining the quark OAM by calculating the integral over phase space of the quark distribution in a longitudinally polarized nucleon multiplied by the naive OAM operator $(\vec b_\perp\times\vec k_\perp)_z$~\cite{OAMYuan}
\begin{equation}\label{lzform}
\ell_z^q=\int\ud x\,\ud^2k_\perp\,\ud^2b_\perp\left(\vec b_\perp\times\vec k_\perp\right)_z\,\rho^{[\gamma^+]q}(\vec b_\perp,\vec k_\perp,x,+\vec e_z).
\end{equation}

\subsection{Overlap representation}

Following the lines of refs.~\cite{Diehl:2000xz,Brodsky:2000xy}, we obtain an overlap representation of eq.~\eqref{lzform} in terms of light-cone wave functions (LCWFs). Since the quark OAM operator is diagonal in light-cone helicity, flavor and color spaces, we write the quark OAM as the sum of the contributions from the $N$-parton Fock states $\ell_z^q=\sum_{N,\beta}\ell_z^{N\beta,q}$, where
\begin{equation}
\ell_z^{N\beta,q}=\frac{i}{2}\int\left[\ud x\right]_N\left[\ud^2k_\perp\right]_N\sum_{i=1}^N\delta_{qq_i}\sum_{n=1}^N\left(\delta_{ni}-x_n\right)
\left[\Psi^{*+}_{N\beta}(r)\left(\vec k_i\times\stackrel{\leftrightarrow}{\nabla}_{k_n}\right)_z\Psi^+_{N\beta}(r)\right].\label{OR}
\end{equation}
%\red{where the integration measures are defined as 
%in ref.~\cite{Lorce:2011kd}}. 
\\
Recently, it has been suggested, on the basis of some quark-model calculations, that the transverse-momentum dependent distribution $h_{1T}^\perp$ may also be related to the quark OAM~\cite{She:2009jq,Avakian:2010br}. It is however important to note that in this case, the quark OAM is defined in more naive terms
\begin{equation}\label{OR2}
\mathcal L_z^{N\beta,q}=-\frac{i}{2}\int\left[\ud x\right]_N\left[\ud^2k_\perp\right]_N\sum_{i=1}^N\delta_{qq_i}\left[\Psi^{*+}_{N\beta}(r)\left(\vec k_i\times\stackrel{\leftrightarrow}{\nabla}_{k_i}\right)_z\Psi^+_{N\beta}(r)\right].
\end{equation}

Finally, according to Ji's sum rule~\cite{Ji:1996ek}, the quark OAM can also be extracted from the following combination of GPDs
\begin{equation}\label{ji-sumrule}
L^q_z=\frac{1}{2}\int^1_{-1}\ud x\left\{x\left[H^q(x,0,0)+E^q(x,0,0)\right]-\tilde H^q(x,0,0)\right\}.
\end{equation}
The overlap representation of the $N$-parton Fock state contribution is then given by
\begin{multline}\label{OR3}
L^{N\beta,q}_z=\frac{1}{2}\int\left[\ud x\right]_N\left[\ud^2k_\perp\right]_N\sum_{i=1}^N\delta_{qq_i}\\
\left\{\left(x_i-\lambda_i\right)\left|\Psi^+_{N\beta}(r)\right|^2+Mx_i\sum_{n=1}^N\left(\delta_{ni}-x_n\right)\left[\Psi^{*+}_{N\beta}(r)\,\frac{\stackrel{\leftrightarrow}{\partial}}{\partial k^x_n}\,\Psi^-_{N\beta}(r)\right]\right\}.
\end{multline}

\subsection{Model results and discussion}

We present in Table \ref{OAMtable} the results from a LCCQM and the light-cone version of the chiral quark-soliton model (LC$\chi$QSM) restricted to the three-quark sector~\cite{Lorce:2011kd}. 

\begin{table}[th!]
\begin{center}
\caption{\footnotesize{The results for quark orbital angular momentum (see eqs.~\eqref{OR}, \eqref{OR2} and \eqref{OR3}) from a LCCQM and the LC$\chi$QSM for $u$-, $d$- and total ($u+d$) quark contributions.}}\label{OAMtable}
\begin{tabular}{@{\quad}c@{\quad}c@{\quad}|@{\quad}c@{\quad}c@{\quad}c@{\quad}|@{\quad}c@{\quad}c@{\quad}c@{\quad}}\hline
\multicolumn{2}{@{\quad}c@{\quad}|@{\quad}}{Model}&\multicolumn{3}{c@{\quad}|@{\quad}}{LCCQM}&\multicolumn{3}{c@{\quad}}{$\chi$QSM}\\
\multicolumn{2}{@{\quad}c@{\quad}|@{\quad}}{$q$}&$u$&$d$&Total&$u$&$d$&Total\\
\hline
$\ell^q_z$&eq.~\eqref{OR}&$0.131$&$-0.005$&$0.126$&$~~0.073$&$-0.004$&$0.069$\\
$\mathcal L^q_z$&eq.~\eqref{OR2}&$0.169$&$-0.042$&$0.126$&$~~0.093$&$-0.023$&$0.069$\\
$L^q_z$&eq.~\eqref{OR3}&$0.071$&$~~0.055$&$0.126$&$-0.008$&$~~0.077$&$0.069$\\
\hline
\end{tabular}
\end{center}
\end{table}

First of all, we see that all definitions give the same values for the \emph{total} quark OAM. This is expected since once summed over all flavors $q$, the expressions \eqref{OR}, \eqref{OR2} and \eqref{OR3} become identical. We see also that there is more net quark OAM in the LCCQM ($\sum_q L^q_z=0.126$) than in the LC$\chi$QSM ($\sum_q L^q_z=0.069$). However, all definitions disagree for the separate flavor contributions. Both the LCCQM and the LC$\chi$QSM predict that $\ell^q_z$ (from the Wigner functions) and $\mathcal L^q_z$ (from the $h_{1T}^\perp$ TMD) are positive for $u$ quarks and negative for $d$ quarks, with the $u$-quark contribution larger than the $d$-quark contribution in absolute value. For $L^q_z$ (from the Ji's sum rule) the LCCQM predicts the same positive sign for the $u$ and $d$ contributions, with the isovector combination $L^u_z-L^d_z>0$, similarly to a variety of relativistic quark model calculations. Instead, the LC$\chi$QSM gives $L^u_z<0$ and $L^d_z>0$, and thereefore $L^u_z-L^d_z<0$, in agreement with lattice calculations~\cite{Hagler:2007xi}.

It is surprising that $\ell^q_z\neq L^q_z$ since it is generally believed that Jaffe-Manohar and Ji's OAM should coincide in absence of gauge degrees of freedom~\cite{Burkardt:2010he}. Note that a similar observation has also been made in the instant-form version of the $\chi$QSM~\cite{Wakamatsu:2005vk}. One may argue that this comes from a problem on the model side (\emph{e.g.} spurious surface contributions or violation of rotational symmetry). This is unlikely considering that the isoscalar part is the same $\ell_z=L_z$. It is only for the separate contributions that the different definitions disagree. Unless rotational symmetry implies non-trivial relations between the different components of the LCWFs, one should expect in general different results for \eqref{OR} and \eqref{OR3}. We emphasize also that the different contributions to OAM depend on the reference point with respect to which OAM is defined. While it is clear that $\ell^q_z$ and $\mathcal L^q_z$ are defined with respect to the transverse center of momentum and origin of axes, respectively, it is not clear from the overlap representation \eqref{OR3} with respect to which point Ji's OAM is defined.

\section{Conclusion}

In summary, we have presented the first model calculation of the Wigner distributions within the light-cone formalism, using a derivation which is not spoiled by relativistic corrections. The results within a light-cone constituent quark model and the light-cone version of the chiral quark-soliton model are very similar, and allowed us to sketch some general features about the behavior of the quarks in the nucleon when observed in the $\vec b_\perp$-plane at fixed $\vec k_\perp$, or in the $\vec k_\perp$-plane at fixed $\vec b_\perp$. We discussed also three different definitions of quark OAM and compared them using the above two models. While all definitions agree on the isoscalar combination, they disagree on the isovector one. Possible reasons for this observation have been discussed.

\acknowledgments
C. L. is thankful to INFN and the Department of Nuclear and Theoretical Physics of the University of Pavia for their hospitality. This work was supported in part by the Research Infrastructure Integrating Activity ``Study of Strongly Interacting Matter'' (acronym HadronPhysics2, Grant Agreement n. 227431) under the Seventh Framework Programme of the European Community, by the Italian MIUR through the PRIN 2008EKLACK ``Structure of the nucleon: transverse momentum, transverse spin and orbital angular momentum''.

\end{document}